# First rare-earth phosphate aerogel: sol-gel synthesis of monolithic ceric hydrogen phosphate aerogel

Yorov K.E.[1], Shekunova T.O.[1,2], Baranchikov A.E.[1,2], Kopitsa G.P.[3,4], Almásy L.[5], Skogareva L.S.[2], Kozik V.V.[6], Malkova A.N.[7], Lermontov S.A.[7], Ivanov V.K.[2,8]*

[1] Lomonosov Moscow State University

[2] Kurnakov Institute of General and Inorganic Chemistry of the Russian Academy of Sciences

[3] Petersburg Nuclear Physics Institute of National Research Centre "Kurchatov Institute"

[4] Grebenshchikov Institute of Silicate Chemistry of the Russian Academy of Sciences

[5] Institute for Solid State Physics and Optics, Wigner Research Centre for Physics, Hungarian Academy of Sciences

[6] National Research Tomsk State University

[7] Institute of Physiologically Active Compounds of the Russian Academy of Sciences

[8] Lomonosov Moscow State University of Fine Chemical Technologies

**Annotation**

Since the late 1960s, ceric hydrogen phosphates have attracted the attention of scientists due to remarkable ion exchange, sorption, proton-conduction and catalytic properties. In this work, through the application of various solvents, we, for the first time, have obtained monolithic aerogels based on ceric hydrogen phosphates with high porosity (~99%) and extremely low density (~10 $\mu g/cm^3$). The composition and structure of aerogels were thoroughly studied with XRD, TEM, SEM, XPS, low temperature nitrogen adsorption methods, TGA/DSC, FTIR and SANS. The aerogels were found to belong to the fibrous macroporous aerogels family.

**1. Introduction**

Aerogels, being highly porous materials with low density and high specific surface area, are gels in which the liquid phase is completely replaced by a gaseous phase [1]. Aerogels are typically used as catalysts, sensors, ion-exchange materials, and as heat and sound insulators [2]. They can be produced from a number of simple substances, both inorganic and organic compounds; hybrid organic-inorganic aerogels are also known [3–17].

The most studied $SiO_2$-based aerogels, obtained by hydrolysis of silicon alkoxides, with subsequent supercritical drying, consist of isotropic $SiO_2$ nanoparticles (0D) forming a spatial network [18]. Recently, a significant number of papers have also been published on the production of aerogels containing anisotropic 1D and 2D building blocks, including carbon nanotubes and graphene [19–22]. Special attention has been paid to monolithic materials having a similar architecture, but containing not only micropores (<2 nm) and mesopores (2–50 nm), but also macropores, since they have a highly accessible surface, which is guaranteed by macroporosity. In addition to the high permeability of such materials, which is important for their application as catalysts and sensors, they can be easily recovered for repetitive use [23].

One of the most challenging tasks in the design of new multifunctional materials is the production of aerogels from orthophosphates of transition and rare-earth elements. Such materials could attract a great deal of interest, due to their extended applications in various possible applications, such as ion-exchange for water purification, catalysts, proton conductors, *etc.* At the same time, the data on the synthesis of such materials is extremely scarce. Thus, methods are reported for the synthesis of aerogels of the compositions $Ti_3(PO_4)_4/Si_3(PO_4)_4$, $AlPO_4/Si_3(PO_4)_4$, $Si_3(PO_4)_4$, $Zn_3(PO_4)_2$, $Zn_3(PO_4)_2/Si_3(PO_4)_4$, $AlPO_4$, $Ti_3(PO_4)_4$ и $Zr_3(PO_4)_4/Si_3(PO_4)_4$ [24]. Zhu *et al.* [23] described the synthesis of a monolithic aerogel based on zirconium phosphate, which can be used for the purification of water as a heavy metals sorbent. Studies concerning the preparation of surface-modified, phosphate-containing oxide aerogels have been also reported [25–28]. In particular, Boyse *et al.* [25] obtained $Nb_2O_5$-based aerogels containing 5 or 10 mol.% of niobium phosphate, which showed a high level of catalytic activity in the reaction of butene-1 isomerisation.

The synthesis of rare-earth phosphate aerogels has not been reported, to date, although they may be of considerable interest, due to a number of valuable properties, including sorption [29], ion exchange [30], proton-conduction [31] and catalytic activities [32], which are typical, in particular, of cerium (IV) hydroorthophosphates. Despite the long history of cerium phosphates [33,34], cerium (III) orthophosphates of monazite or rhabdophane structures [35] remain the most investigated, whereas cerium (IV) orthophosphates have been studied to a much lesser degree. Only recently, crystalline structures of two acidic phosphates of cerium (IV), $Ce(PO_4)(HPO_4)_{0.5}(H_2O)_{0.5}$ and $Ce(PO_4)_{1.5}(H_2O)(H_3O)_{0.5}(H_2O)_{0.5}$ were solved by Nazaraly *et al.* [36–38].

In the present study, we made the first attempt to obtain aerogels based on ceric hydrogen phosphates, to yield materials possessing higher porosity, and better mechanical strength and chemical stability, in comparison to wet gels. Guidance for our attempt was the fact that wet gels of monolithic ceric hydrogen phosphates are easily formed by mixing cerium-containing hydrogen phosphate solutions with water [39]. It was reasonable to assume that, under certain conditions, these gels could be dried supercritically, resulting in monolithic aerogel materials.

## 2. Materials and methods

The following materials were used, as received and without further purification: $Ce(NO_3)_3 \cdot 6H_2O$ (99%, Aldrich #238538), orthophosphoric acid (85 wt.% aq., analytical grade, Khimmed Russia), aqueous ammonia (25 wt.%, extra-pure grade, Khimmed Russia), isopropanol (extra-pure grade, Khimmed Russia), acetonitrile (analytical grade, Khimmed Russia), methyl *tert*-butyl ether (Acros, 99%), distilled or deionised (18 MΩ) water.

To obtain cerium (IV) orthophosphate gels, we used a technique developed in our group recently [40]. A typical procedure for the synthesis of a monolithic wet gel involves the dissolution of 0.020 g of $CeO_2$ powder (prepared according to [41]) in 1 ml of $H_3PO_4$ (85 wt.%, $\rho$ = 1.689 g/cm$^3$) under constant stirring at 100°C. After the $CeO_2$ was completely dissolved and the solution showed no Tyndall effect, the solution was cooled to room temperature, (the solution is hereafter referred to as **sCeP**, molar ratio Ce:P is equal to 1:140), and 4 ml of distilled water or 4 ml of methyl *tert*-butyl ether (MTBE) was added. As a result, a gel was formed, which was aged for 10 days, then the solvent was replaced by keeping the gel in acetonitrile for one week, with a daily solvent change. The samples were dried under supercritical conditions, to obtain the aerogels. As solvents for the supercritical drying, we used carbon dioxide and MTBE. A glass tube containing wet gel under an MTBE layer (14–16 mL) was placed in an autoclave (V = 38 mL). The autoclave was mounted in a furnace, heated at a rate of 100°C/h to 235–245°C (6.0–7.0 MPa) and held at that temperature for 10–15 min. Next, the pressure in the heated autoclave was gradually lowered to atmospheric pressure and the autoclave was evacuated for 30 min (30–40 kPa), cooled and opened.

Supercritical drying in $CO_2$ was carried out in a system that comprised a Supercritical 24 high pressure pump for $CO_2$ (SSI, USA), a 50 mL steel reactor and a BPR back pressure regulator (Goregulator, Waters, USA). The sample was washed with liquid $CO_2$ for 2 h at a temperature of 20°C and pressure of 15 MPa. The temperature in the reactor was then raised to 50°C and the sample was washed with supercritical $CO_2$ (15 MPa) for 2–2.5 h. Next, the pressure in the heated autoclave was gradually (30–40 min) lowered to atmospheric pressure and the autoclave was cooled and opened [42].

As a reference sample, a xerogel was used, the synthesis of which included mixing the cerium-containing phosphoric acid solution and water, purifying the resulting gel by dialysis against deionised water and drying the purified gel at 60°C under atmospheric pressure.

X-ray powder diffraction patterns were recorded with a Bruker D8 Advance diffractometer using CuK$_\alpha$ radiation in the 2θ range 3–120° at a 2θ step of 0.02° and a counting time of 0.3 s per step.

The microstructure of the samples was studied by means of transmission electron microscopy (TEM) with a Leo912 AB Omega analytical transmission electron microscope. TEM images were taken at an accelerating voltage of 100 kV in the bright-field mode.

The microstructure (scanning electron microscopy, SEM) and the chemical composition (energy dispersive X-ray analysis, EDX) of the samples were analysed on a Carl Zeiss NVision 40 high-resolution scanning electron microscope equipped with an Oxford Instruments X-MAX (80 mm$^2$) detector, operating at an accelerating voltage of 1–20 kV. SEM images were taken with an Everhart-Thornley detector (SE2) at 1 kV accelerating voltage.

The investigation of the chemical composition of the surface layers of ceric hydrogen phosphate materials was conducted by X-ray photoelectron spectroscopy (XPS) on a SPECS X-ray photoelectron spectrometer with a PHOIBOS-150 energy analyser in fixed analyser transmission mode (15 eV), using MgKα radiation ($hv$ = 1253.6 eV). Experimental data processing was performed with the CasaXPS software package.

Specific surface areas of powders were determined by low temperature nitrogen adsorption on an ATX-6 analyser (Katakon, Russia), using the 5-point Brunauer–Emmett–Teller (BET) model at partial pressures in the range 0.05–0.25. Pore size distribution was assessed by the Barrett–Joyner–Halenda (BJH) method, using adsorption isotherms at partial pressures in the range 0.4–0.97.

Thermal analysis was performed on a TGA/DSC/DTA SDT Q-600 analyser (TA Instruments), upon linear heating to 1,000°C (heating rate of 10°C/min) in a 250 ml/min airflow.

The FTIR spectra of the samples were recorded on a Perkin Elmer Spectrum One spectrometer, in a range of 450–4000 cm$^{-1}$, in attenuated total reflectance mode.

The SANS experiment was performed using the "Yellow Submarine" instrument of the BNC research reactor, in Budapest (Hungary), which operates in near point geometry. The use of two neutron wavelengths ($\lambda$ = 4.9 and 9.4 Å, $\Delta\lambda/\lambda$ = 18%) and two sample-to-detector distances (1.57 and 5.5 m) provided measurements in the momentum transfer range of $6\cdot10^{-3} < q < 0.3$ Å$^{-1}$. The BerSANS software [43] was used for data pre-processing.

## 3. Results and discussion

Directly upon addition of distilled water to the sCeP solution, the formation of a solid phase (gel) was observed. A stable and cohesive gel was formed at the ratios $V_{sCeP} : V_{water}$ = 1:2 – 1:6 (see Table 1). When a smaller volume of water was added, the gel was not formed; with the ratio of $V_{sCeP} : V_{water} > 1:8$, the resulting gel broke down after 1 week of ageing.

Table 1. Formation of ceric phosphate hydrogels at different volume ratios of reactants

| Entry | Volume of sCeP solution, (ml) | Water volume, (ml) | Molar ratio Ce:H$_2$O | Formation of gel | Ageing for 7 days |
|---|---|---|---|---|---|
| V1 | 1 | 1 | 1:530 | No | – |
| V2 | 1 | 2 | 1:1060 | Yes | Gel remains monolithic |
| V4 | 1 | 4 | 1:2120 | Yes | Gel remains monolithic |
| V6 | 1 | 6 | 1:3180 | Yes | Gel remains monolithic |
| V8 | 1 | 8 | 1:4250 | Yes | Gel breaks down |
| V10 | 1 | 10 | 1: 5300 | Yes | Gel breaks down |

Interestingly, the formation of the gel was observed not only when the cerium-containing phosphate solution interacted with distilled water, but also when sCeP was mixed with non-aqueous solutions, including protic or aprotic solvents (methanol, ethanol, isopropanol, ethylene glycol, acetone, methyl *tert*-butyl ether and tetrahydrofuran). Table 2 contains the results of the corresponding experiments. The reason for the observed effect is still unclear, and, for its clarification, data are required on the ceric coordination chemistry in non-aqueous orthophosphoric acid solutions, which are rather scarce. It should be noted that the obtained gels did not always remain monolithic upon ageing (Table 2, experiments S2, S5, S9, S10). The breakdown of gels in these cases was probably associated with the reduction of Ce$^{+4}$ ions to Ce$^{+3}$, due to the oxidation of the corresponding solvent under acidic (H$_3$PO$_4$) conditions. To determine which of the obtained gels remain monolithic after solvent replacement, the obtained samples were further dialysed against deionised water (see Table 2). The results of this experiment revealed that only in entries S1 and S8 (see Table 2) did the samples remain monolithic.

Table 2. The effect of solvent nature on the formation of Ce (IV) phosphate gel

| Entry | Solvent | Formation of gel | After aging (12 days), when transferred to a test tube with distilled water | After replacing the solvent with distilled water |
|---|---|---|---|---|
| S1 | Water | Yes | Gel remains monolithic | Gel remains monolithic |
| S2 | Methanol | Yes | Gel disintegrates | – |
| S3 | Ethanol | Yes | Gel remains monolithic | Gel disintegrates |
| S4 | Isopropanol | Yes | Gel remains monolithic | Gel disintegrates |
| S5 | Acetone | Yes | Gel disintegrates | – |
| S6 | Acetonitrile | No | – | – |
| S7 | Diethyl ether | No | – | – |
| S8 | MTBE | Yes | Gel remains monolithic | Gel remains monolithic |
| S9 | Tetrahydrofuran | Yes | Gel disintegrates | – |
| S10 | Ethylene glycol | Yes | Gel disintegrates | – |
| S11 | Hexane | No | – | – |

Hence, in the following, for the synthesis of aerogels, we used wet gels obtained by adding the distilled water or MTBE to a solution of cerium-containing phosphoric acid in a volume ratio of 4:1 (see Table 1, Table 2, experiments S1, S8). The selected gels, upon replacement of the solvent with acetonitrile, were dried under supercritical conditions in $CO_2$ and in MTBE (Fig. 1). For the sake of clarity, the samples synthesised in this way are hereafter named as follows:

| **Notation** | **CePH** | **CePM** | **CePMTB** | **CePx*** |
|---|---|---|---|---|
| **Solvent used for wet-gel synthesis** | $H_2O$ | MTBE | MTBE | $H_2O$ |
| **Solvent used for supercritical drying** | $CO_2$ | $CO_2$ | MTBE | ambient pressure drying in air |

* xerogel obtained as a reference sample according to the reported synthetic procedure [40]

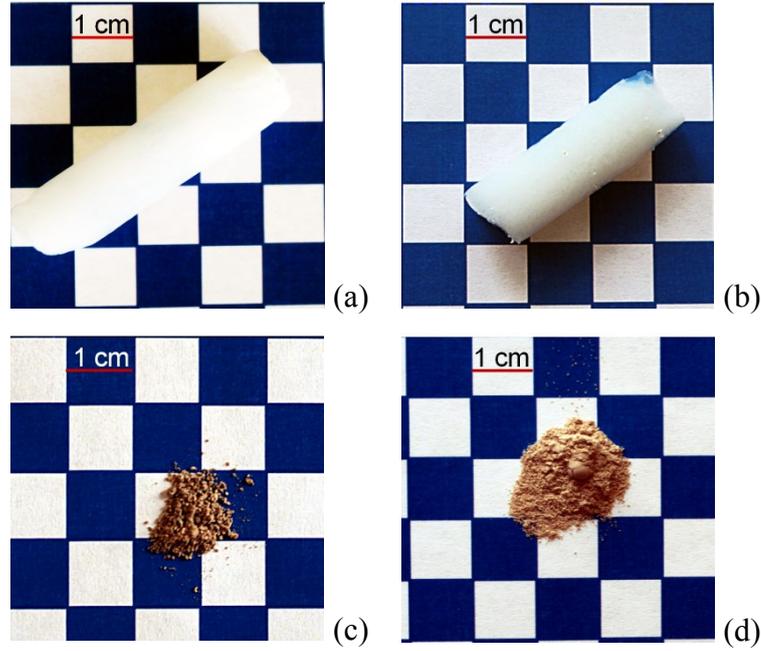

Fig. 1. The appearance of the obtained samples: (a) – CePH, (b) – CePM, (c) – CePMTB, (d) – CePx.

It was demonstrated that supercritical drying in $CO_2$ leads to the successful production of monolithic aerogels (Fig. 1a, b). Their mechanical strength is quite high for common handling, but somewhat lower than that of silica aerogels synthesized by common procedures. Note that monolithic aerogel samples are visually white, whereas powdered samples prepared from the same hydrogels are coloured (Fig. 1). The diffuse UV-vis reflectance spectra (Fig. S1) of colourless CePH and coloured CePx samples are nearly identical, indicating that the aerogels' colourlessness is probably due to their low density and multiple scattering of ambient light by the solid-gas interfaces. Actually, the geometrical density of monolithic aerogels (CePH and CePM) amounts to ~0.01 g/cm$^3$. The porosity of the obtained aerogels was estimated using the formula [44]:

$$P = \left(1 - \frac{\rho_1}{\rho_2}\right) \cdot 100\%,$$

where $\rho_1$ is the aerogel density and $\rho_2$ is the skeletal density. The skeletal density of aerogel samples, as measured using a Pycnomatic ATC helium pycnometer, was equal to 2.98±0.04 g/cm$^{-3}$. The calculated porosity of the monolithic aerogels was ~ 99%.

According to X-ray diffraction analysis of aerogels obtained by drying in supercritical $CO_2$ and xerogel, CePx are mostly X-ray amorphous (Fig. 2 a, b, c).

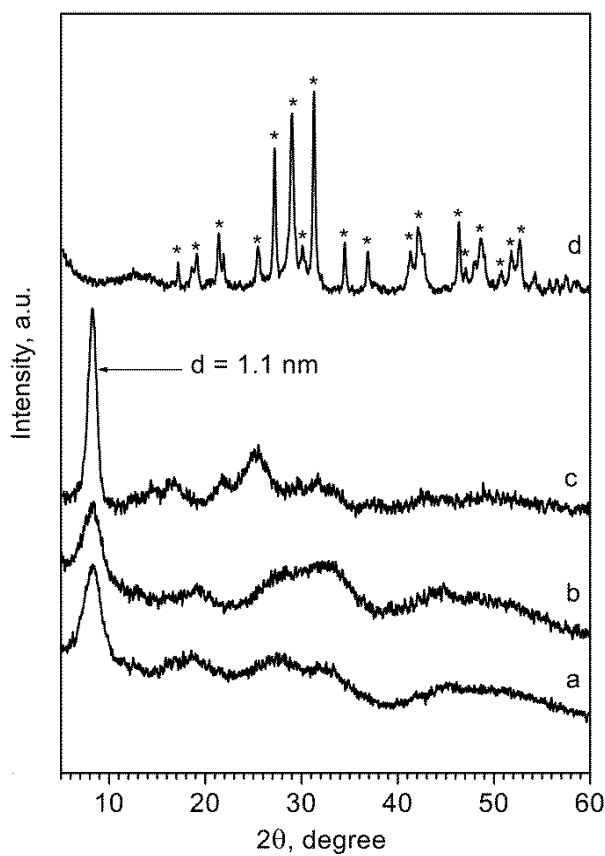

Fig. 2. X-ray diffraction patterns of ceric hydrogen phosphate aerogels and xerogels: (a) CePx, (b) CePH, (c) CePM, (d) CePMTB. The peaks corresponding to the monazite phase are indicated with an asterisk.

In the diffractograms of these samples, at 2θ ~8°, one can observe a pronounced broadened maximum, which may indicate the existence of short-range order in cerium-containing hydrogen orthophosphate gels with a characteristic distance of ~1.1 nm. The presence of this peak for cerium-phosphate gels has also been reported previously [40,45]. Hayashi et al. [46] presumed that the presence of this peak is due to the layered structure of the gels allowing them to be intercalated with e.g. oleylamine [46]. The diffractogram for the CePM aerogel (Fig. 2c) is almost the same as the diffractogram for the Ce (IV) hydrogen phosphate of $Ce(HPO_4)_2 \cdot 3.5H_2O$ composition, which was obtained by Hayashi *et al.* by a procedure involving the preparation of $Ce(SO_4)_2 \cdot 4H_2O$ solution in 0.5 M $H_2SO_4$, followed by its addition to a solution of 6M $H_3PO_4$ (95°C), followed by digestion at 95°C [46].

The CePMTB sample (Fig. 2 d) was a single-phase monazite (CePO$_4$, PDF2 №32-199). The formation of trivalent cerium orthophosphate seems to be the result of the oxidation of methyl *tert*-butyl ether during synthesis at high temperature (about 250°C), (probably with the formation of *tert*-butyl formate and *tert*-butyl alcohol), with simultaneous reduction of Ce (IV) [47]. As supercritical drying in the MTBE medium failed to yield a monolithic aerogel, further detailed studies of the CePMTB sample were not conducted.

Fig. 3 shows the results of thermal analysis of CePx, CePM and CePH samples. In general, both xerogel (CePx) and aerogels (CePH and CePM) exhibited similar behaviour upon heating, and their thermal decomposition was of a multistage nature. In the region of relatively low temperatures (up to ~200°C), physically bound molecules of water, acetonitrile and others were removed. At temperatures up to ~500°C, apparently, the removal of chemically bound water occurred [40,48,49].

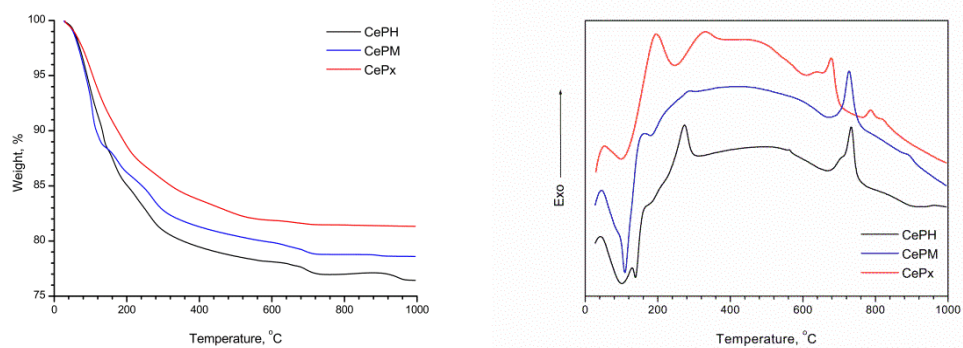

Fig. 3. Thermogravimetric and differential thermal analysis data for ceric hydrogen ophosphate aerogels (CePH, CePM) and xerogel CePx.

At 680–750°C, a pronounced exothermic effect was observed, accompanied by a small weight loss (0.5–1%). This effect has previously been described in the literature [40,50], and is associated with the decomposition of Ce (IV) orthophosphates, with the release of oxygen and the formation of crystalline Ce (III) phosphate. According to XRD data, this led to the formation of monazite phase (CePO$_4$, PDF №00-032-199) and cerium tripolyphosphate (CeP$_3$O$_9$, PDF №00-033-0336). It is noteworthy that the position of this exothermic effect when heating the aerogels (CePM and CePH) shifted to higher temperatures in comparison with the corresponding parameter for xerogel (CePx). Most likely, this difference was due to the differences in chemical composition of the corresponding samples, namely to the higher phosphate to cerium molar ratio in aerogels.

Based on the results of a local EDX analysis (Table 3), the average Ce:P molar ratio was established as being approximately 1:1.5 for CePx xerogel and 1:2 for CePH and CePM aerogels. Thus, the chemical composition of CePx xerogel is close to Ce$_2$H(PO$_4$)$_3$·H$_2$O [51],

Ce$_2$(PO$_4$)$_2$HPO$_4$(H$_2$O) [51] and Ce(PO$_4$)$_{1.5}$(H$_2$O)(H$_3$O)$_{0.5}$(H$_2$O)$_{0.5}$ [37], while the composition of CePH and CePM aerogels is closer to compounds Ce(HPO$_4$)$_2$·xH$_2$O [52] and CeO(H$_2$PO$_4$)$_2$·2H$_2$O [46]. Fig. 4 provides the survey XPS spectra for CePH and CePM aerogels, as well as for initial nanocrystalline CeO$_2$. In Table 3, the molar ratios of the elements are provided, which were calculated from the integral intensities of the corresponding signals, taking into account the element sensitivity coefficients. The P/Ce ratios for aerogels computed from XPS data were in good agreement with the EDX data (Table 3).

Table 3. Ce:P molar ratios in the samples of CePM and CePH aerogels and CePx xerogel, according to EDX and XPS

| Sample | P:Ce ratio (EDX) | P:Ce ratio (XPS) | O/(Ce+P) ratio (XPS) |
|---|---|---|---|
| CePx | 1.5 | – | – |
| CePH | 2.0 | 1.9 | 2.4 |
| CePM | 2.1 | 2.3 | 2.5 |

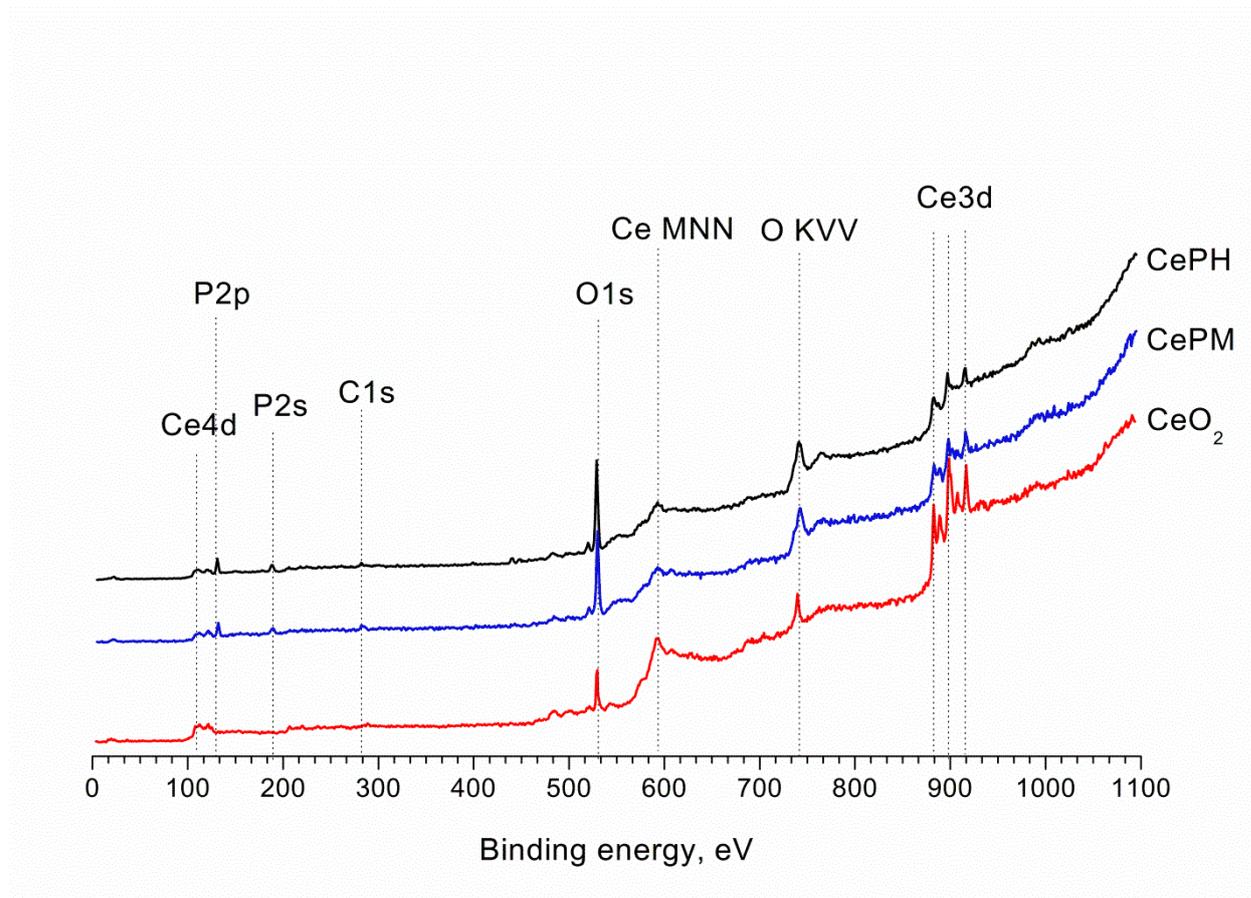

Fig. 4. XPS survey spectra of nano-CeO$_2$ and CePH, CePM aerogels.

Fig. 5 shows the IR spectra of aerogel and xerogel samples. The general appearance of the IR spectra of CePx, CePH and CePM samples is virtually the same, and is typical for the rare-earth metal phosphates. The absorption maxima at 1,061 (v.str), 980 (br), 620 (str) and 540 (med) cm$^{-1}$ (see Fig. 5a, b) correspond to the four characteristic symmetric and antisymmetric oscillations of the P-O-bonds in the $PO_4^{3-}$ ion. The absence of the pronounced splitting of the absorption bands of $PO_4$ groups in the IR spectra in the regions of 900-1,200 cm$^{-1}$ and 600 cm$^{-1}$ may indicate that the orthophosphate anion in the structure of the obtained substances did not exhibit definite denticity, and could act both as a monodentate, and as a polydentate, ligand. The absorption maxima with the wave numbers 3,225 (str) and 1,628 (str) cm$^{-1}$ characterise the stretching vibrations of the OH groups and deformation vibrations of H-O-H in water molecules, respectively. The band at 1,435 cm$^{-1}$ refers to the valence vibrations of $CO_3^{2-}$, and indicates the presence of a carbonate impurity, (the sample was air-dried). The band at 1,240 cm$^{-1}$ for the CePM sample can be attributed to the C-O-C oscillations typical of MTBE [53], but, since it was present in the IR spectra of other samples, it is more likely to have been attributed to the P–O–H plane deformation [50].

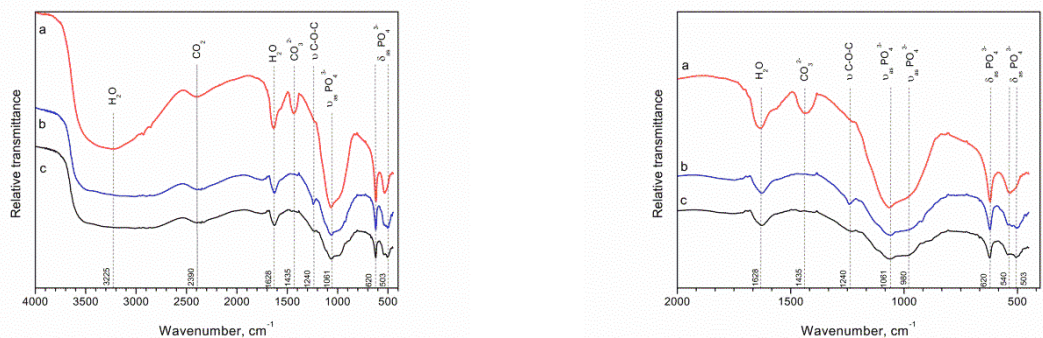

Fig. 5. IR spectra of ceric hydrogen phosphate xerogel CePx (a), and ceric hydrogen phosphate aerogels CePM (b) and CePH (c).

Table 4 shows the specific surface areas of the samples. The relatively low specific surface area (60-70 m$^2$/g) and high porosity of aerogels (~99%) indicate the presence of a large number of macropores in aerogels.

Table 4. Specific surface area of ceric hydrogen orthophosphate aerogel and xerogel samples.

| Sample | CePx | CePH | CePM |
|---|---|---|---|
| $S_{sp}$, m$^2$/g | 60 ± 6 | 70 ± 7 | 75 ± 8 |

Fig. 6 shows the full nitrogen adsorption/desorption isotherm for the CePM aerogel. One can see that the isotherm has a narrow hysteresis, which can be attributed to H3 type according to IUPAC classification [54]. The isotherm reflects the presence of macropores as evidenced by the continuous uptake at the highest relative pressure range. Unfortunately, the method of low-temperature nitrogen adsorption is not sensitive to pores larger than ~100 nm [55].

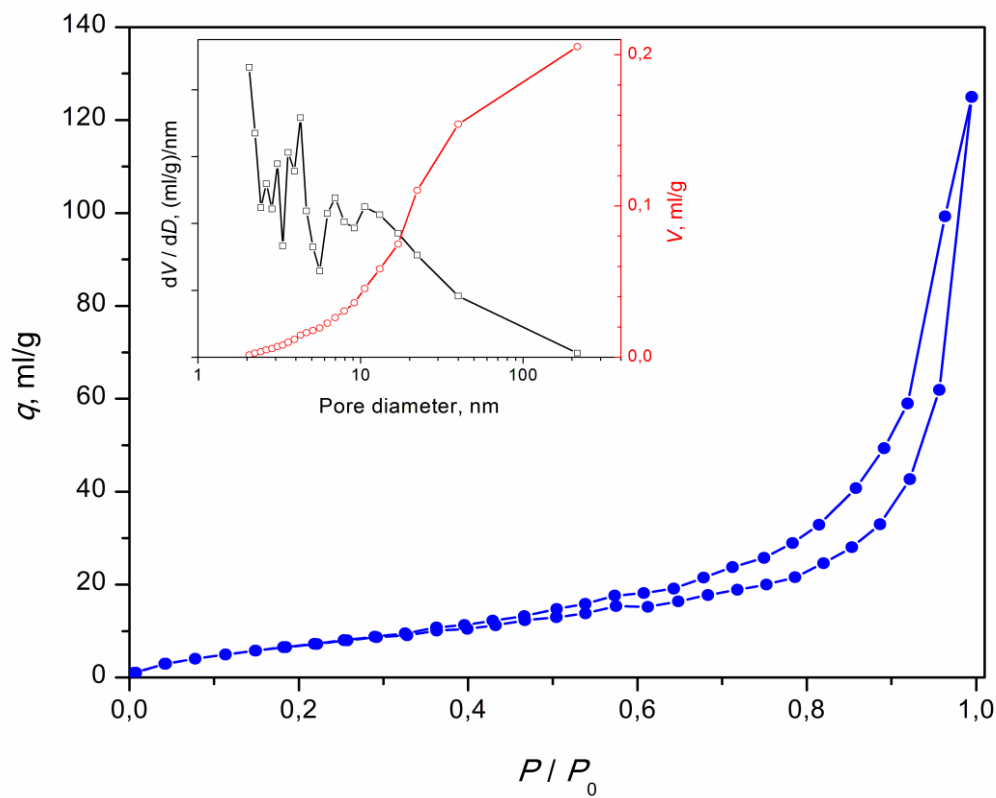

Fig. 6. Nitrogen adsorption/desorption isotherm, pore size distribution and differential pore volume distribution (pore volume density) (in the inset) for the CePM aerogel.

Fig. 7 shows TEM images of CePx, CePH and CePM samples. The results of transmission electron microscopy and electron diffraction indicated that CePx, CePH and CePM samples consisted of fibre-shaped anisotropic particles, at that the smallest fibre thickness (about 15–20 nm) was observed for the CePx sample (Fig. S2). The fibre diameter for the CePH and CePM samples was significantly (approximately 2 times) larger, and, according to the TEM data, they contained inhomogeneities, most probably the closed pores (Figs. S2, 7a, b).

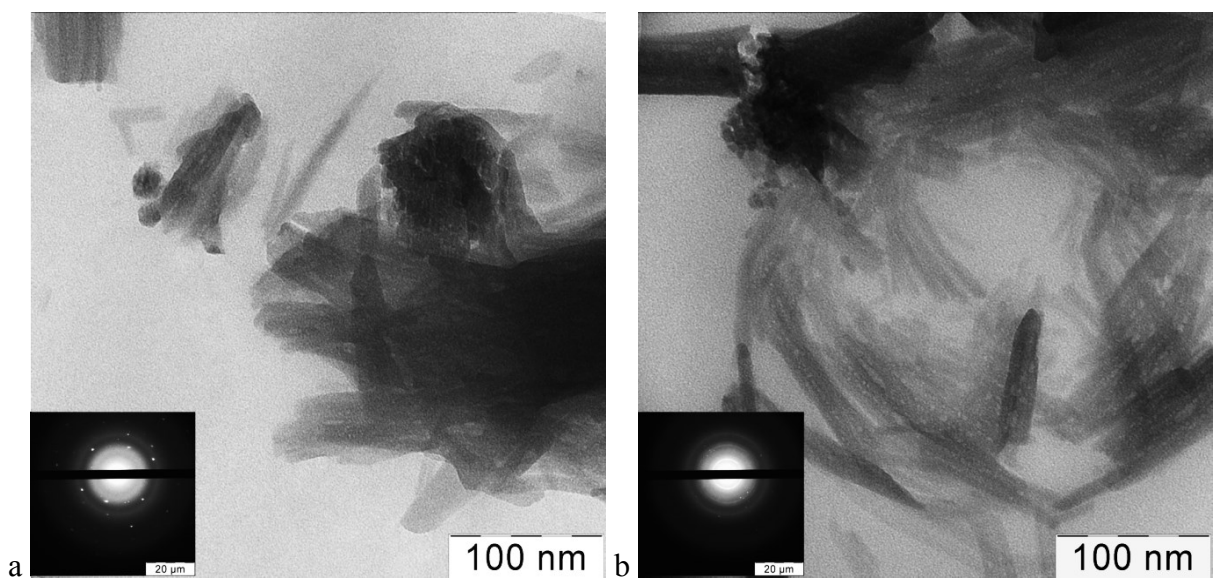

Fig. 7. TEM images of ceric hydrogen phosphate aerogels CePH (a) and CePM (b).

The results of scanning electron microscopy (Fig. 8) were in good agreement with the transmission electron microscopy data, and also indicated that CePH and CePM aerogels are characterised by a larger fibre diameter (up to 40-50 nm), compared to CePx xerogel. SEM data also indicated a slight decrease in fibre length (up to 1-2 microns) in aerogels, compared to CePx xerogel (Figs. S3, 8). In addition, the obtained SEM images confirmed the presence of a significant number of macropores in aerogels, with pores from 100 to 500 nm.

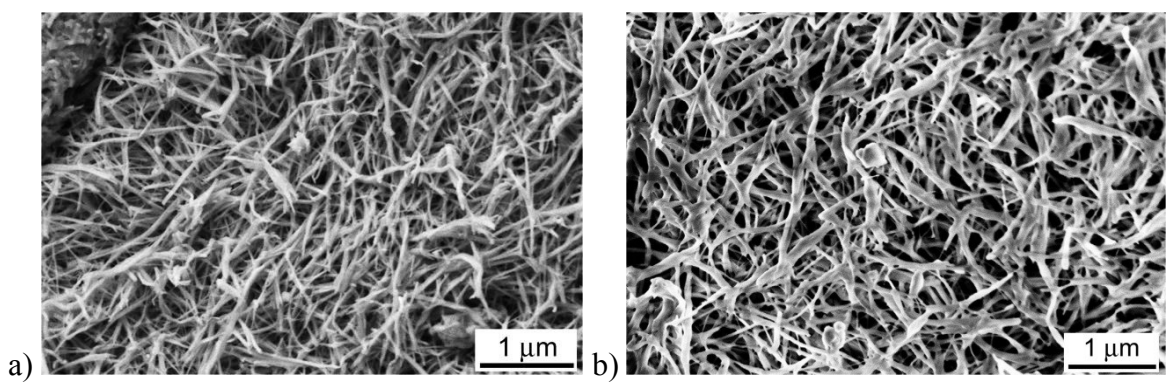

Fig. 8. SEM images of ceric hydrogen phosphate aerogels CePM (a) and CePH (b).

The structure of xerogel (CePx) and aerogel (CePM) samples was independently analysed by means of small angle neutron scattering. Experimental curves of the differential macroscopic neutron cross section $d\Sigma(q)/d\Omega$ versus momentum transfer $q$ (Fig. 9) evidently showed three characteristic $q$-ranges where the behaviours of the SANS cross section $d\Sigma(q)/d\Omega$ were significantly different.

In the range $0.02 < q < 0.15$ Å$^{-1}$, the scattering cross section $d\Sigma(q)/d\Omega$ for all the samples satisfied the power law $q^{-n}$. The exponent $n$ values found from the slope of the straight-line sections of the experimental curves for CePx and CePM samples were equal to 3.32 ± 0.02 and 3.96 ± 0.02, respectively. The *n* value for the CePx sample corresponds to the scattering from the fractal surface with the dimension $D_s = 6 - n = 2.68 \pm 0.02$. The *n* value for the aerogel sample was very close to 4, which corresponds to the scattering on inhomogeneities with virtually smooth (in the scale of neutron wavelength used in the experiment, $\lambda = 4.9$ Å) surfaces ($D_s = 2.04 \pm 0.02$) according to Porod law. Such a difference in fractal properties of xerogel and aerogel samples obtained from the nearly identical starting materials is quite unusual. Probably, surface fractal structure of the xerogel is formed upon dehydration under ambient conditions. Fractalization of smooth materials upon their drying has been observed previously (see e.g. [56]).

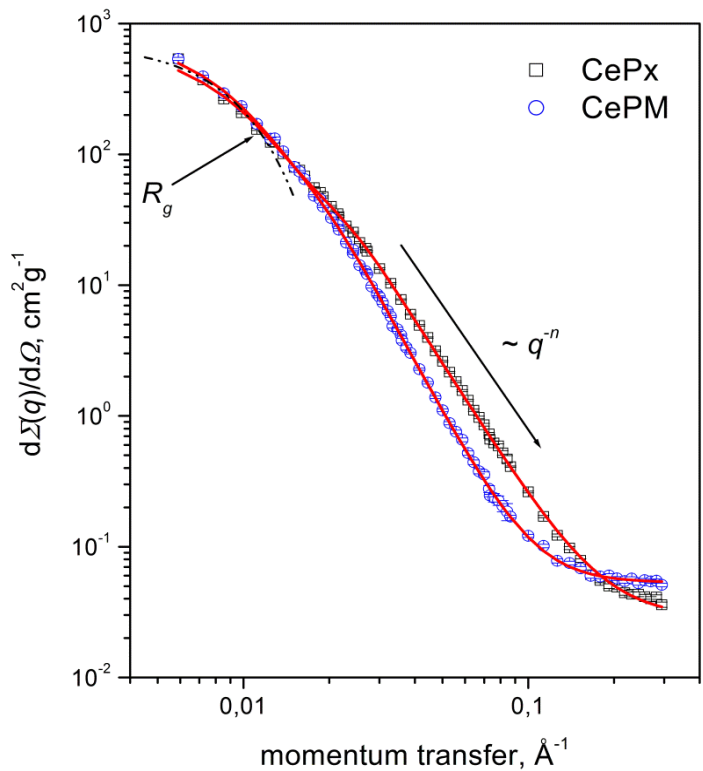

Fig. 9. SANS differential cross section $d\Sigma(q)/d\Omega$ for the samples of ceric hydrogen phosphate xerogel CePx and aerogel CePM. Fitting of experimental data was performed according to [57] and is shown as solid lines.

Deviations from the power law $q^{-n}$ in differential cross section curves $d\Sigma(q)/d\Omega$ were observed for all samples, both at small and large *q*-ranges. The deviation in the small *q*-range ($q < 2 \cdot 10^{-2}$ Å$^{-1}$) was due to approaching the Guinier regime [58], where the scattering is governed by the inhomogeneities having the characteristic size $R_c$. Analysis of the scattering in this *q*-range enables an estimation of the $R_c$ value, which is equal to 25.9±0.6 nm for CePx, and 29.8±0.7 nm for CePM samples. These values correspond to the upper estimate of the fibrillae radii in the gels and are in line with TEM and SEM data. In the large $q$-range ($q > 0.15$ Å$^{-1}$), the cross-sections d$\Sigma(q)$/d$\Omega$ became constant, (did not depend on $q$), because of incoherent scattering on hydrogen atoms present in the samples in the form of physically and chemically bound water, and scattering on the inhomogeneities having a size comparable to the neutron wavelength used in the experiment.

    The combination of the obtained data on the micro- and mesostructure of the obtained aerogels makes it possible to classify them as macroporous materials composed of interlaced fibres. Such macroporous aerogels are typically obtained using specific approaches, such as template synthesis [23], freeze-drying synthesis [59,60] and supercritical drying with rapid removal of supercritical fluid [60]. The macroporous structure is also inherent to aerogels consisting of fibrous or lamellar particles with carbon nanotube-based and cellulose-based aerogels, (which can be referred to as "1D-aerogels"), as well as graphene ("2D aerogels") being classic examples of such materials. It is also reasonable to mention the unusual class of alumina 1D aerogels obtained by the controlled oxidation of metallic aluminum on the surface of aluminum amalgam [61], or the hydrolysis of aluminum trichloride in the presence of propylene oxide [3]. Jung *et al.* [62] proposed a method for producing monolithic macroporous aerogels using both one-dimensional (nanorods of Ag, Si, MnO$_2$ and single-walled nanotubes) and two-dimensional nanoparticles (MoS$_2$, h-BN and graphene). The method is based on the self-assembly of the preliminary anisotropic 1D or 2D nanoparticles obtained into a cross-linking network upon the increase in the concentration of their colloidal suspensions. In turn, direct sol-gel methods for the synthesis of 1D aerogels of complex inorganic compounds are practically unknown. This is because the formation of flexible fibres that could form an interwoven, stable network within a sol-gel process is not typical for such structures. One of the scarce examples of sol-gel derived 1D aerogels is vanadium pentoxide gels derived from neutral VO(OH)$_3$ species or vanadium alkoxide precursors [63]. These gels form according to polymeric growth mechanism which can be judged as one of the possible mechanisms for cerium phosphate fibrillae formation, too. The formation of gels from poorly soluble complex inorganic compounds, in particular orthophosphates of transition and rare-earth elements, is also complicated, because of their low solubility in the liquid phase, which leads to their hasty precipitation in the form of fine precipitates, hindering the structuration to impart the desirable morphology.

**Conclusions**

Monolithic aerogels based on ceric hydrogen orthophosphates, being the first representatives of the 1D rare-earth phosphate aerogel family, have been obtained and characterised. It has been shown that these gels have a specific surface of 60–70 m$^2$/g and, in addition, have a high porosity (~99%), due to the presence of pores with a wide range of sizes, including macropores. The obtained aerogels had a fibrous structure, with a fibre thickness up to ~50 nm and a length of 1–2 μm, which allows considering them as a new type of 1D aerogel based on complex inorganic compounds.

**Acknowledgements**

We are grateful to Dr. O.V. Boytsova and Dr. N.P. Simonenko for their kind assistance. The effect of the supercritical fluid on the aerogels' structure and composition was conducted with the support of the Russian Science Foundation (14-13-01150). This research was performed using the equipment of the JRC PMR IGIC RAS.

# Supplementary material

# First rare-earth phosphate aerogel: sol-gel synthesis of monolithic ceric hydrogen phosphate aerogel

Yorov K.E.[1], Shekunova T.O.[1,2], Baranchikov A.E.[1,2], Kopitsa G.P.[3,4], Almásy L.[5], Skogareva L.S.[2], Kozik V.V.[6], Malkova A.N.[7], Lermontov S.A.[7], Ivanov V.K.[2,8]*

[1] Lomonosov Moscow State University

[2] Kurnakov Institute of General and Inorganic Chemistry of the Russian Academy of Sciences

[3] Petersburg Nuclear Physics Institute of National Research Centre "Kurchatov Institute"

[4] Grebenshchikov Institute of Silicate Chemistry of the Russian Academy of Sciences

[5] Institute for Solid State Physics and Optics, Wigner Research Centre for Physics, Hungarian Academy of Sciences

[6] National Research Tomsk State University

[7] Institute of Physiologically Active Compounds of the Russian Academy of Sciences

[8] Lomonosov Moscow State University of Fine Chemical Technologies

* van@igic.ras.ru

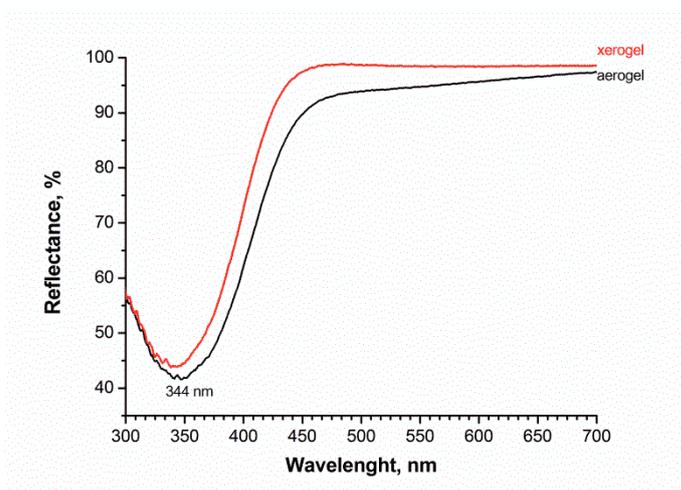

Fig. S1. Diffuse UV-vis reflectance spectra of CePx (xerogel) and CePH (aerogel) samples.

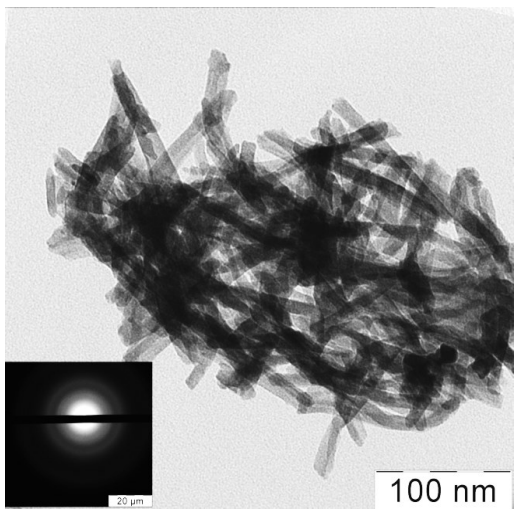 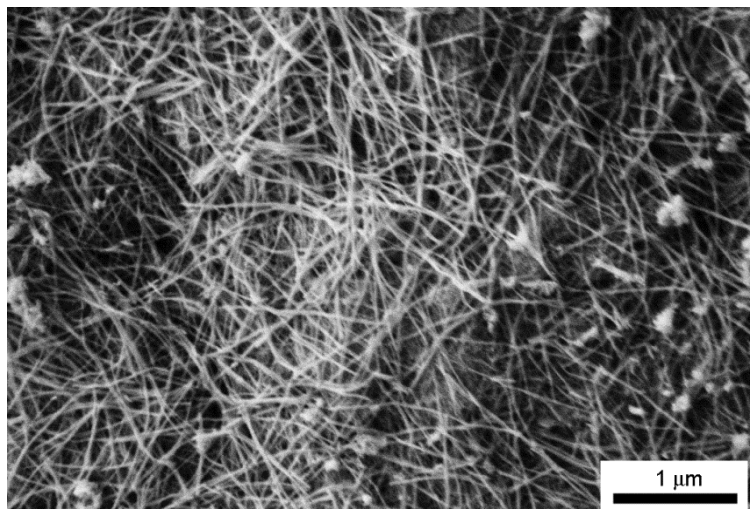

Fig. S2. TEM images of ceric hydrogen phosphate xerogel CePx.

Fig. S3. SEM images of ceric hydrogen phosphate xerogel CePx.